\documentclass[twocolumn,showpacs,preprintnumbers,amsmath,amssymb,aps]{revtex4}
\usepackage{graphicx}
\usepackage{dcolumn}
\usepackage{bm}
\usepackage{color}
\begin{document}
\preprint{}
\title{`$f_0(600)$' and chiral dynamics}
\author{T. K. Jha{\footnote {email: tkjha@prl.res.in}}}
\affiliation 
{
Theoretical Physics Division, Physical Research Laboratory, Navrangpura, 
Ahmedabad, India - 380 009
}
\date{\today}
\begin{abstract}
The role of light scalar meson `$f_0 (600)$' is investigated in nuclear matter in an Effective chiral model in the mean-field approach. For the purpose, we scan the properties of the matter at various saturation densities imposing constraint such as the vacuum value of the pion decay constant $f_{\pi} \approx 131 ~MeV$. With a simple approach, the bound on the mass of the scalar meson is calculated and found in the range $m_{\sigma} = 546 \pm 10 MeV$. Further, the present analysis show that nuclear matter favor high nucleon effective mass and dominant repulsive forces at high density, the insight and the implications of which are discussed.
\end{abstract}
\maketitle
\section{Introduction}

Quantum Hadro-dynamics (QHD) is an effective tool to understand nuclear force with baryons and mesons as the degree of freedom to provide a microscopic description of finite nuclei or infinite nuclear matter \cite{wal74}. The well known renormalizable quantum field theory was introduced by Walecka based on baryons and a scalar and vector meson to study high density matter in the mean field theory, where the meson field operators are replaced by their vacuum expectation value or are treated as the classical fields. The model known as the $\sigma-\omega$ model \cite{wal74} and its extensions (complemented with cubic and quartic non-linearities of the $\sigma$ meson) \cite{boguta77} have been phenomenal in this respect. Recently, there have been some efforts to extend the models including various interaction terms and generate new parameter sets in the relativistic mean field formalism \cite{rmf,rmf1,rmf2} with motivation to extend the applicability of the model in nuclear matter studies. Parallel to that, another class of model, the sigma model evolved with the required features for a relativistic theory with chiral symmetry, the symmetry of the QCD or strong interactions. Originally introduced by Gell-Mann and Levy \cite{ch01}, later Lee \& Wick \cite{le74} emphasized the importance of chiral symmetry in nuclear matter studies. Another interesting aspect was the introduction of dynamically generated isoscalar vector meson mass, first introduced by Boguta \cite{ch03} to obtain the saturation properties of nuclear matter. It is now well established that nuclear matter saturation is realized through a balance between the scalar (attractive) and the vector (repulsive) component of the nuclear force mediated via mesons. Among them, the mesons that are known to be important are the $\pi(140)$, $\sigma(400 - 1200)$, $\omega(783)$ and the $\rho(770)$ meson. As evident, the mass of the scalar meson ($\sigma$) is not precisely known which seems to invoke challenge and interest for both the experimentalist and the theorist on account of its importance in nuclear and particle physics. The estimate from the Particle Data Group (PDG) quotes the mass of this scalar meson `$f_0 (600)$' or $\sigma-$meson in the range $(400 - 1200)$ MeV \cite{pdg}, seen as a $\pi\pi$ resonance. Since they are fundamental and constitute the Higgs sector of the strong interaction, it is important and inevitable in the description of fundamental properties of matter and hence commands special attention.

The range of the interaction and thus the fundamental properties of nuclear matter is sensitive to the mass of the $\sigma-$meson. For example, the inclusion of higher order interaction terms of the scalar field is known to considerably reduce the nuclear incompressibility, which is a fundamental constant of nature. Apart from that there are important astrophysical implications related to nuclear incompressibility. With regard to chiral symmetry, the $\sigma-$meson plays the role of the chiral partner of pion and the symmetry is spontaneously broken in its ground state, as a consequence of which the scalar condensate attains a finite expectation value and the pions are rendered massless. The value of the scalar condensate reflects the strength with which the symmetry is broken and experimentally, this value is found to be $\approx$ 131 MeV \cite{pdg} as the pion decay constant `$f_{\pi}$'. The rather precise value of the pion decay constant can serve as an alternative way to provide theoretical limits to the scalar meson mass. In the present work, we look into these aspects and analyze the correlations between properties such as nucleon effective mass ($m^{\star}$), the nuclear incompressibility ($K$), the scalar meson mass ($m_{\sigma}$) with varying saturation density ($\rho_0$).

What follows in the next section is the description of the model attributes and the methodology adopted for the aforesaid purpose. In section III, we present the results and the analysis of the present work. A comparison of our results with other theoretical and the experimental findings is also presented. Finally we conclude with some important findings of the present work. 

\section{The Effective Chiral Model}

The model we consider for the present analysis embodies dynamically generated mass of the vector meson. The explicit dependence of such a mass term on the scalar condensate then regulates the value of the nucleon effective mass and leads to the saturation of nuclear matter. Using sigma model with dynamically generated mass for vector meson, Glendening studied finite temperature aspects of nuclear matter and its application to neutron stars \cite{gl}, but without $\rho-$meson and its isospin symmetry influence on matter. Although a nice framework respecting chiral symmetry, a drawback was its unacceptable high incompressibility and in the subsequent extension of the model \cite{five}, the mass of the vector meson is not generated dynamically. The model that we consider \cite{tkj08} in our present analysis embodies higher orders of the scalar field in addition to the dynamically generated mass of the vector meson. Here, we intend to analyze in detail the attributes of the model with respect to the inherent vacuum properties of chiral symmetry and its effect on the resulting EOS. This would in turn interlink various properties of matter with the scalar condensate. We now proceed to briefly describe the salient features of the present model.

The effective Lagrangian of the model interacting through the exchange of the pseudo-scalar meson $\pi$, the scalar meson $\sigma$, the vector meson $\omega$ and the iso-vector $\rho-$meson is given by \cite{tkj08}:

\begin{eqnarray}
\label{lag}
{\cal L}&=& \bar\psi_{B}~\left[ \big(i\gamma_\mu\partial^\mu
         - g_{\omega}\gamma_\mu\omega^\mu
        - \frac{1}{2}g_{\rho}{\vec \rho}_\mu\cdot{\vec \tau}
          \gamma^\mu\big )\right] 
\nonumber \\
        &-& \bar\psi_{B}~\left[g_{\sigma~}~\big(\sigma + i\gamma_5
             \vec \tau\cdot\vec \pi \big)\right]~ \psi_{B}
\nonumber \\
        &+& \frac{1}{2}\big(\partial_\mu\vec \pi\cdot\partial^\mu\vec\pi
        + \partial_{\mu} \sigma \partial^{\mu} \sigma\big)
\nonumber \\
        &-& \frac{\lambda}{4}\big(x^2 - x^2_0\big)^2
        - \frac{\lambda b}{6 m^2}\big(x^2 - x^2_0\big)^3
        - \frac{\lambda c}{8 m^4}\big(x^2 - x^2_0\big)^4
\nonumber \\
        &-& \frac{1}{4} F_{\mu\nu} F_{\mu\nu}
        + \frac{1}{2}{g_{\omega B}}^{2}x^2 \omega_{\mu}\omega^{\mu}
\nonumber \\
        &-& \frac {1}{4}{\vec R}_{\mu\nu}\cdot{\vec R}^{\mu\nu}
        + \frac{1}{2}m^2_{\rho}{\vec \rho}_{\mu}\cdot{\vec \rho}^{\mu}\ .
\end{eqnarray}

The first line of the above Lagrangian represents the interaction of the nucleon isospin doublet $\psi_B$ with the aforesaid mesons. In the second line we have the kinetic term followed by the non-linear terms in the pseudo-scalar-isovector pion field `$\vec \pi$', the scalar field `$\sigma$', and higher order terms of the scalar field in terms of the invariant combination of the two i.e., $x^2= {\vec \pi}^2+\sigma^{2}$. Finally in the last two lines, we have the field strength and the mass term for the vector field `$\omega$' and the iso-vector field `$\vec \rho$' meson. $g_{\sigma}, g_{\omega}$ and $g_{\rho}$ are the usual meson-nucleon coupling strength of the scalar, vector and the iso-vector fields respectively. Here we shall be concerned only with the normal non-pion condensed state of matter, so we take $<\vec \pi>=0$ and also $m_{\pi} = 0$.

The interaction of the scalar and the pseudoscalar mesons with the vector boson generates a dynamical mass for the vector bosons through spontaneous breaking of the chiral symmetry with scalar field attaining the vacuum expectation value $x_0$. Then the mass of the nucleon ($m$), the scalar ($m_{\sigma}$) and the vector meson mass ($m_{\omega}$), are related to $x_0$ through

\begin{eqnarray}
m = g_{\sigma} x_0,~~ m_{\sigma} = \sqrt{2\lambda} x_0,~~
m_{\omega} = g_{\omega} x_0\ .
\end{eqnarray}
\noindent

In the mean-field ansatz, the vector field ($\omega$), the scalar field ($\sigma$) (in terms of $Y=x/x_0 = m^{\star}/m$) and the isovector field ($\rho$) is respectively given by

\begin{eqnarray}
\omega_0=\sum_{B}\frac{ \rho_B }{g_{\omega} x^2},
\end{eqnarray}

\begin{eqnarray}
(1-Y^2) &-& \frac{b}{m^2 c_{\omega}}(1-Y^2)^2 +\frac{c}{m^4c_{\omega}^2}(1-Y^2)^3 
\nonumber \\
&+& \frac{2 c_{\sigma}c_{\omega}\rho_B^2}{m^2Y^4} -\frac{2 c_{\sigma}\rho_S}{m Y}=0\,
\label{effmass}
\end{eqnarray}

\begin{equation}
\rho_{03} =\sum_{B} \frac{g_{\rho}}{m_\rho^2} I_{3}~\rho_{B}.
\end{equation}

The quantity $\rho_B$ and $\rho_S$ are the vector and the scalar density defined as,
\begin{equation}
\rho_B= \frac{\gamma}{(2\pi)^3}\int^{k_F}_o d^3k, \hskip 0.2in \rho_{S}= \frac{\gamma}{(2\pi)^3}\int^{k_F}_o\frac{m^{\star} d^3k}
         {\sqrt {k^2+m^{\star 2}}}.
\end{equation}

\noindent
In the above, `$k_F$' is the fermi momenta of the baryon and $\gamma=4$ (symmetric matter) is the spin degeneracy factor. For symmetric nuclear matter ($N = Z$), we neglect the contribution from the $\rho-$meson. The nucleon effective mass is then $m^{\star} \equiv Y m$ and $c_{\sigma}\equiv  g_{\sigma}^2/m_{\sigma}^2 $ are $c_{\omega} \equiv g_{\omega}^2 /m_{\omega}^2 $ are the scalar and vector coupling constants that enters as a parameter in our calculations. From the expression for the scalar field equation (eqn. 4), which computes the nucleon effective mass ($m^{\star}$ in terms of Y = m/$m^{\star}$), it can be seen that the there is an explicit dependence of the vector field contribution (fourth term). The total energy density `$\varepsilon$' and pressure `$P$' of symmetric nuclear matter for a given baryon density is then given by the following.

\begin{eqnarray}
\varepsilon
&=&
	  \frac{\gamma}{2\pi^2} \int _o^{k_F} k^2dk\sqrt{{k}^2 + m^{\star 2}}
        + \frac{m^2(1-Y^2)^2}{8c_{\sigma}} \nonumber \\
        &-& \frac{b}{12c_{\omega}c_{\sigma}}(1-Y^2)^3
        + \frac{c}{16m^2c_{\omega}^2c_{\sigma}}(1-Y^2)^4 \nonumber\\
        &+& \frac{c_{\omega} \rho_B^2}{2Y^2}
\\
P &=&	\frac{\gamma }{6\pi^2} \int _o^{k_F} \frac{k^4dk}{\sqrt{{k}^2 + m^{\star 2}}}
        - \frac{m^2(1-Y^2)^2}{8c_{\sigma}} \nonumber \\
        &+& \frac{b}{12c_{\omega}c_{\sigma}}(1-Y^2)^3
        - \frac{c}{16m^2c_{\omega}^2c_{\sigma}}(1-Y^2)^4 \nonumber\\
        &+& \frac{c_{\omega}\rho_B^2}{2Y^2}
\end{eqnarray}

The meson field equations are solved self-consistently at a fixed baryon density and the corresponding energy density and pressure is calculated. We need to evaluate the parameters of the model (the coupling constants $C_{\sigma}$, $C_{\omega}$, $C_{\rho}$ and the higher order scalar field constants $B$ and $C$) that satisfy nuclear matter saturation properties, which we describe next. 

The individual contribution to the total energy density for symmetric nuclear matter can be abbreviated as

\begin{eqnarray}
\varepsilon=\varepsilon_k+\varepsilon_{\sigma}+\varepsilon_{\omega},
\end{eqnarray}

where,
\begin{eqnarray}
\varepsilon_k=\frac{\gamma}{2\pi^2}\int _o^{k_F} k^2dk\sqrt{{k}^2 
+ m^{\star 2}}\ , \hskip 0.2in \varepsilon_{\omega}=\frac{c_{\omega} \rho_B^2}{2Y^2},
\end{eqnarray}

and

\begin{eqnarray}
\varepsilon_{\sigma}&=&\frac{m^2(1-Y^2)^2}{8c_{\sigma}} - \frac{b}{12c_{\omega}c_{\sigma}}(1-Y^2)^3 \nonumber \\
&+& \frac{c}{16m^2c_{\omega}^2c_{\sigma}}(1-Y^2)^4.
\end{eqnarray}

\noindent
In the above expressions, $\rho_B = \rho_n + \rho_p$ is the total baryon density which is the sum of the neutron density `$\rho_n$' and the proton density `$\rho_p$'. The relative neutron excess is then given by $\delta = (\rho_n - \rho_p)/\rho_B$. At the standard state $\rho_B = \rho_0$, the nuclear matter saturation density and $\delta = 0$. Consequently, at the standard state ($\rho_0, 0$), the energy per particle is $e (\rho_0, 0)$ = $\varepsilon/\rho_0$ - m = $a_1$ $\approx$ -16 MeV for symmetric nuclear matter and hence,

\begin{eqnarray}
\varepsilon = \varepsilon_k+\varepsilon_{\sigma}+\varepsilon_{\omega}=\rho_0(m-a_1).
\end{eqnarray}

At the equilibrium condition $P (\rho_0, 0)=0$, we have,

\begin{eqnarray}
P &=& -\varepsilon~+~ \rho_B~\frac{\partial \varepsilon}{\partial \rho_B}\nonumber \\
&=& \frac{1}{3} \varepsilon_k ~-~ \frac{1}{3} m^{\star} \rho_S ~-~ \varepsilon_{\sigma} ~+~ \varepsilon_{\omega} ~=~ 0.
\end{eqnarray}

Using eqn. (14) and eqn. (15), the respective energy contributions can be expressed in terms of the properties at the saturation density as, 

\begin{eqnarray}
\varepsilon_{\sigma}=\frac{1}{2} ~\left[\rho_0(m-a_1)-\frac{1}{3}
(2\varepsilon_k+m^{\star}\rho_s)\right]
\end{eqnarray}
\noindent
and

\begin{eqnarray}
\varepsilon_{\omega}=\frac{1}{2}\left[\rho_0(m-a_1)-\frac{1}{3}
(4\varepsilon_k-m^{\star}\rho_s)\right],
\end{eqnarray}

where $\rho_s$, the scalar density is given by,

\begin{eqnarray}
\rho_s=\frac{1}{\pi^2}m^{\star} \left[ k_F E_F 
- ln \Big(\frac{k_F+E_F}{m^{\star}}\Big) m^{\star 2}\right].
\end{eqnarray}

In the above equations, $m^{\star}=Ym$ is the nucleon effective mass and $E_F=\sqrt{k_F^2+m^{\star 2}}$ is the effective energy of the nucleon carrying momenta $k_F$. From eqn. (13), the vector coupling ($C_{\omega}$) can be readily evaluated using the relation

\begin{eqnarray}
C_{\omega}=\frac{2 Y^2}{\rho_0^2} \varepsilon_{\omega},
\end{eqnarray}
\noindent
with $\varepsilon_{\omega}$ given by eqn. (17), for a specified value of $Y=m^{\star}/m$ defined at $\rho_0$. Similarly, the scalar coupling can be calculated using the relation 

\begin{eqnarray}
C_{\sigma} &=& \frac{m Y}{2 \rho_S} \left[(1-Y^2) - \frac{b}{m^2 c_{\omega}}(1-Y^2)^2 + \frac{c}{m^4c_{\omega}^2}(1-Y^2)^3\right] \nonumber \\
&+& \frac{m Y}{2 \rho_S} \left[\frac{2 c_{\sigma}c_{\omega}\rho_B^2}{m^2Y^4}\right].
\end{eqnarray}

The higher order scalar field couplings constants `b' and `c' can be evaluated simultaneously using the relations,

\begin{eqnarray}
b &=& \frac{6 c_{\sigma} c_{\omega} m \rho_S}{Y (1-Y^2)^2} 
+ \frac{6 c_{\sigma} c_{\omega}^2 \rho_B^2}{Y^4 (1-Y^2)^2} \nonumber \\
&-& \frac{48 \varepsilon_{\sigma} c_{\sigma} c_{\omega}}{(1-Y^2)^3} + \frac{3 m^3 c_{\omega}}{(1-Y^2)},
\end{eqnarray}

\begin{eqnarray}
c&=& \frac{8 c_{\sigma} c_{\omega}^2 m^3 \rho_s}{Y (1-Y^2)^3}
- \frac{8 c_{\sigma} c_{\omega}^3 m^2 \rho_B^2}{Y^4 (1-Y^2)^3}\nonumber \\
&-& \frac{48 \varepsilon_{\sigma} c_{\sigma} c_{\omega}^2}{(1-Y^2)^4} + \frac{2 c_{\omega}^2 m^4}{(1-Y^2)^2}.
\end{eqnarray}

For asymmetric matter, the $\rho-$ meson coupling ($C_{\rho}$) can be fixed for a desirable value of $J = 32 MeV$ as,

\begin{equation}
J = \frac{c_{\rho} k_F^3}{12\pi^2} + \frac{k_F^2}{6\sqrt{(k_F^2 + m^{\star 2})}}\ ,
\end{equation}
where $c_{\rho} \equiv g^2_\rho/m^2_{\rho}$ and $k_F=(6\pi^2\rho_B/\gamma)
^{1/3}$.

The nuclear incompressibility at saturation density is given as,

\begin{equation}
K = 9~\rho_0^2~\frac{\partial^2 (\varepsilon/\rho_B)}{\partial \rho_B} \Big|_0.
\end{equation}

For a detailed methodology to evaluate the model parameters to the saturation properties of symmetric nuclear matter, in the mean-field approach one can refer to \cite{cons09,param1,param2}.

\section{Results and discussions}

\begin{table*}
\caption{For non-linear (1st row) and the linear (2nd row) interactions in the scalar field tabulated are the model parameters that satisfies the nuclear matter saturation properties such as binding energy per nucleon ($B/A - m$), the nucleon effective mass $Y = m^{\star}/m $ and the asymmetry energy coefficient $J \approx 32$ MeV for saturation densities ranging from $\rho_0 = (0.12 - 0.18) fm^{-3}$. The nucleon, the vector meson and the isovector vector meson masses are taken to be 939 MeV, 783 MeV and 763 MeV respectively and $c_{\sigma} = (g_{\sigma}/ m_{\sigma})^2$, $c_{\omega} = (g_{\omega}/ m_{\omega})^2$ and $c_{\rho} = (g_{\rho}/ m_{\rho})^2$ are the coupling constants for the scalar, vector and isovector fields. $B = b/m^2$ and $C = c/ m^4$ are the higher order constants of the scalar field interactions and $B,C=0$ for the linear model. Other derived quantities such as the scalar meson mass `$m_{\sigma}$', the pion decay constant `$f_{\pi}$' and the nuclear matter incompressibility ($K$) are given along with the nuclear radius constant ($r_0$).}
\begin{center}
\begin{tabular}{ccccccccccccccccccc}
\hline
\hline
\multicolumn{1}{c}{set}&
\multicolumn{1}{c}{$c_{\sigma}$}&
\multicolumn{1}{c}{$c_{\omega}$} &
\multicolumn{1}{c}{$c_{\rho}$} &
\multicolumn{1}{c}{$B$} &
\multicolumn{1}{c}{$C$} &
\multicolumn{1}{c}{$\rho_0$} &
\multicolumn{1}{c}{$\varepsilon /\rho_B - M$} &
\multicolumn{1}{c}{$m_{\sigma}$} &
\multicolumn{1}{c}{$Y$} &
\multicolumn{1}{c}{$f_{\pi}$} &
\multicolumn{1}{c}{$K$} &
\multicolumn{1}{c}{$r_0$} \\
\multicolumn{1}{c}{ } &
\multicolumn{1}{c}{($fm^2$)} &
\multicolumn{1}{c}{($fm^2$)} &
\multicolumn{1}{c}{($fm^2$)} &
\multicolumn{1}{c}{($fm^2$)} &
\multicolumn{1}{c}{($fm^4$)}&
\multicolumn{1}{c}{($fm^{-3}$)}&
\multicolumn{1}{c}{($MeV$)}&
\multicolumn{1}{c}{($MeV$)} &
\multicolumn{1}{c}{($m^{\star}/m$)} &
\multicolumn{1}{c}{($MeV$)} &
\multicolumn{1}{c}{($MeV$)} &
\multicolumn{1}{c}{($fm^{-1}$)} \\
\hline
1  &8.730 &2.261 &6.309 &-8.030 &0.544 &0.120 &-16.3 &477.86 &0.871 &131.233 &231 &1.257 \\
   &7.574 &3.777 &6.151 &--  &-- &-- &-16.0 &663.11 &0.790 &101.529 &644 &-- \\
2  &8.270 &2.278 &6.129 &-6.834 &0.384 &0.125 &-16.3 &492.81 &0.865 &130.738 &252 &1.241 \\
   &7.267 &3.602 &5.979 &--  &-- &-- &-15.8 &661.03 &0.790 &103.982 &633 &-- \\
3  &7.863 &2.272 &5.959 &-5.834 &0.085 &0.130 &-16.3 &504.80 &0.860 &130.899 &270 &1.224 \\
   &7.010 &3.439 &5.812 &--  &-- &-- &-16.0 &657.72 &0.789 &106.402 &650 &-- \\
4  &7.436 &2.257 &5.794 &-4.709 &2.497 &0.135 &-16.3 &517.32 &0.855 &131.350 &287 &1.209 \\
   &6.811 &3.314 &5.502 &--  &-- &-- &-16.2 &654.95 &0.787 &108.402 &650 &-- \\
5  &7.209 &2.243 &5.636 &-4.372 &0.933 &0.140 &-16.3 &523.74 &0.850 &131.761 &303 &1.194 \\
   &6.588 &3.171 &5.502 &--  &-- &-- &-16.0 &652.92 &0.787 &110.813 &645 &-- \\
6  &6.918 &2.265 &5.486 &-3.659 &0.678 &0.145 &-16.3 &537.31 &0.843 &131.111 &326 &1.181 \\
   &6.424 &3.077 &5.358 &--  &-- &-- &-16.3 &649.92 &0.784 &112.486 &674 &-- \\
7  &6.688 &2.264 &5.341 &-3.213 &0.093 &0.150 &-16.3 &546.32 &0.837 &131.142 &354 &1.167 \\
   &6.205 &2.955 &5.224 &-- &-- &-- &-16.2 &647.99 &0.784 &114.794 &665 &-- \\
8  &6.446 &2.261 &5.202 &-2.696 &0.155 &0.155 &-16.3 &556.10 &0.831 &131.241 &378 &1.155 \\
   &6.121 &2.892 &5.087 &--  &-- &-- &-16.4 &645.46 &0.780 &116.026 &706 &-- \\
9  &6.206 &2.266 &5.063 &-2.165 &0.265 &0.160 &-16.3 &567.46 &0.824 &131.070 &398 &1.142 \\
   &5.922 &2.782 &4.962 &--  &-- &-- &-16.3 &643.59 &0.780 &118.304 &705 &-- \\
10 &6.016 &2.259 &4.936 &-1.819 &0.110 &0.165 &-16.3 &575.40 &0.818 &131.291 &433 &1.131 \\
   &5.742 &2.682 &4.847 &--  &-- &-- &-16.1 &641.80 &0.780 &120.484 &697 &-- \\
11 &5.765 &2.267 &4.800 &-1.232 &0.704 &0.170 &-16.3 &588.83 & 0.810 &131.064 &480 &1.120 \\
   &5.566 &2.586 &4.733 &--  &-- &-- &-16.0 &639.96 &0.780 &122.723 &693 &-- \\
12 &5.522 &2.268 &4.683 &-0.653 &1.660 &0.175 &-16.3 &601.75 &0.803 &131.039 &530 &1.110 \\
   &5.406 &2.4976 &4.627 &-- &-- &-- &-15.8 &638.19 &0.780 &124.872 &685 &-- \\
13 &5.373 &2.268 &4.556 &-0.441 &1.122 &0.180 &-16.3 &610.03 &0.795 &131.036 &580 &1.100 \\
   &5.251 &2.412 &4.523 &--  &-- &-- &-15.6 &636.40 &0.780 &127.053 &680 &-- \\
\hline
\hline
\end{tabular}
\end{center}
\end{table*}

The spontaneous breaking of chiral symmetry lends mass to the hadrons and relates them to the vacuum expectation value (VEV) of the scalar field ($x_0$), which is what is shown in eqn. (2). Immediately, what follows from the third term in eqn. (2) is that, the VEV of the scalar field which has a minimum potential at $f_{\pi}$ and is directly related to the vector coupling constant $C_{\omega}$ through $x_0 = f_{\pi} = m_{\omega}/g_{\omega} = 1/ \sqrt{C_{\omega}}$. It can therefore be seen that the vector coupling is explicitly constrained from the vacuum value of the pion decay constant. Similarly, a little rearrangement reveal that the mass of the scalar meson can be given by  $m_{\sigma} = m~{\sqrt C_{\omega}/{\sqrt C_{\sigma}}}$. In order to evaluate the parameters of the model we demand that the matter is bounded from below, i.e., the coefficient `C' in the quartic scalar field term remains positive and the resulting energy per particle for symmetric nuclear matter is $\approx -16 MeV$.

With the methodology described in the previous section, we evaluated the nuclear matter parameters of the model with both the linear ($L\sigma$) and non-linear ($NL\sigma$) interactions in the scalar field. Such an analysis not only enable us to study the underlying differences arising due to scalar field interactions but would also lineate the quantitative contribution of the higher order scalar field to the EOS. For a better correlation, the model parameters are evaluated at various saturation densities ranging from $\rho_0 = (0.12 - 0.18) fm^{-3}$. For linear model, we followed similar procedure but neglected the contributions of the higher order scalar field and so the corresponding constants $B,C = 0$, and is given in the second row of Table I for the specified saturation density $\rho_0$.

\begin{figure}[ht]
\begin{center}
\includegraphics[width=8cm]{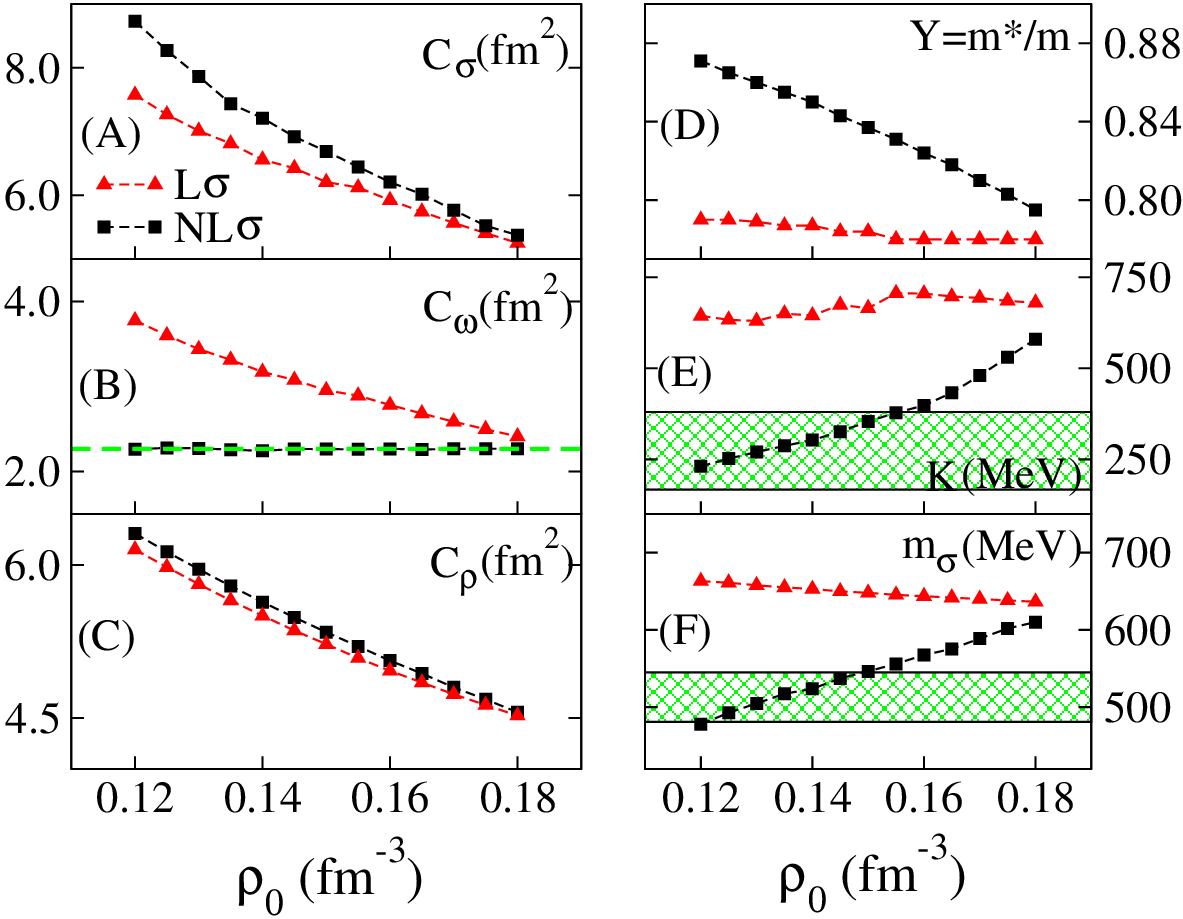}
\end{center}
\caption{The evaluated nuclear matter parameters for the model with non-linear scalar field interaction ($NL\sigma$) and the linear interaction ($L\sigma$) as a function of $\rho_0$ are shown for the following. (A) The scalar coupling, (B) the vector coupling where the green dashed horizontal line indicates the vector coupling strength needed to satisfy the experimental value of the pion decay constant `$f_{\pi}=131 MeV$' (C) the isoscalar vector coupling, which satisfy the constraint $J=32 MeV$ (D) the nucleon effective mass `$Y = m^{\star}/m$', (E) the nuclear matter incompressibility, where the shaded region depicts the bound obtained from HIC data \cite{data02} and (F) the $\sigma-$meson mass, where the shaded region depicts the bound from recent measurement of $m_{\sigma}$ \cite{mura02}.}
\end{figure}

In Fig. 1 (A-F), what is shown is the variation of the coupling constants and various derived physical quantities pertaining to the saturation properties of nuclear matter as a function of saturation density in the present model. From Fig. 1(A), we find that the scalar coupling strength $C_{\sigma}=(g_{\sigma}/m_{\sigma})^2$ differs for the linear and non-linear scalar field interactions, but the difference gradually cease to exist for matter saturating at higher densities, say at $\rho_0 \approx 0.18 fm^{-3}$. Although the couplings in both the case decrease with increase in saturation density. In the case of vector coupling $C_{\omega}=(g_{\omega}/m_{\omega})^2$ shown in Fig. 1(B), this difference is much larger quantitatively but they follow similar trend. As discussed in previous section, the vector coupling strength is directly controlled by the vacuum value of the pion decay constant $f_{\pi}=\sigma_0$, through the spontaneous breaking of the chiral symmetry following eqn. (2). In order that $f_{\pi} = 131 MeV$, we require $C_{\omega} = 2.26 fm^2$ which is shown by the blue dashed horizontal line in Fig. 1(B). It is to be noted here that the non-linearity in the scalar field interactions enables us to satisfy the vacuum value of the pion decay constant at each value of saturation density as evident from the plot. In case of linear interactions, we find it difficult to reproduce $f_{\pi}$ and demand a minimum energy bound for symmetric nuclear matter simultaneously. However for matter saturating at higher density, the underlying differences between the linear and the non-linear interactions dissolve. The case is different for the iso-vector coupling constant `$C_{\rho}$', as it depends on the momenta `$k_f$' and the effective mass of the nucleon ($m^{\star}$), both of which remains nearly same in magnitude for both the linear and the non-linear cases and hence not much of a difference is seen, as evident from Fig. 1(C).

From Fig. 1(D), it is evident that the matter favor higher nucleon effective mass at lower saturation density and vice versa, when non-linear interactions are included. In case of linear model, at the specified range of saturation density, $m^{\star}\approx 0.78 m$, whereas for non-linear model, the nucleon effective mass is predicted in the range $m^{\star} = (0.79 - 0.87)m$. With regard to the present model, the resulting nucleon effective mass in the nuclear medium is a consequence of the interplay of both the scalar and the vector forces as evident from Eqn. (4). The dynamically generated mass of the vector meson plays significant role in the determination of the nucleon effective mass and regulates the scalar coupling strength. With increasing density the vector forces becomes dominant and has a net effect resulting in higher nucleon effective mass. It is to be noted here that the nucleon effective mass at the saturation densities in the specified range is much higher in comparison to other relativistic field theoretical prescriptions found in the literature. We predict that within the preferred range of saturation density $\rho_0 = 0.150 \pm 0.005 fm^{-3}$, the nucleon mass drops by $\approx 18$ \% , which is nearly half the value predicted by NL3 parameterization from the relativistic mean-field theory ($m^{\star}/m = 0.60$ at $\rho_0 = 0.148 fm^{-3}$) \cite{nl3}. However, our prediction is consistent with the values obtained from recent analysis of neutron scattering off lead nuclei ($m^{\star}/m = (0.80 - 0.90)$) \cite{compact,nuclei}. It is worth to recall that a lower nucleon effective mass is known to reproduce the finite nuclei properties, such as the spin-orbit effects splitting correctly \cite{furn98}. 

In Fig. 1(E), the resulting incompressibility of the equation of state at various $\rho_0$ is plotted. It is evident that nuclear matter becomes less compressible or stiffer at higher saturation densities. In comparison with the incompressibility bound inferred from heavy ion collision experiment (HIC) $K = (167 - 380) MeV$ \cite{data02}, we find that the EOS with lower nucleon effective mass or conversely higher saturation density is ruled out. Equivalently, the agreement with the experimental flow data in the density range $2 < \rho_B/\rho_0 < 4.6$ seem to favor repulsion (higher effective mass). Further the HIC data (shaded region in Fig. 1(E)) seems to favor a lower saturation density, i.e., $\rho_0 < 0.155 fm^{-3}$. It is to be noted that the linear parameterization of the model results in stiffer EOS, i.e., EOS with higher value of incompressibility. For example the EOS for $\rho_0 = 0.155 fm^{-3}$ the nonlinear field in the scalar sector accounts for the decrease in incompressibility of about $50 \%$. This softening effect is more pronounced at lower saturation densities and vice versa. For example for $\rho_0 = 0.12 fm^{-3}$, the incompressibility is lowered by $\approx 65 \%$, where as it lowers by $\approx 15 \%$ in case of $\rho_0 = 0.18 fm^{-3}$. Incompressibility is a fundamental constant of nature and is also the most important quantity for supernova explosions and neutron stars \cite{k1}, which dictates the balance between gravity and internal pressure of the stellar system.

\begin{figure}[ht]
\begin{center}
\includegraphics[width=3.8cm]{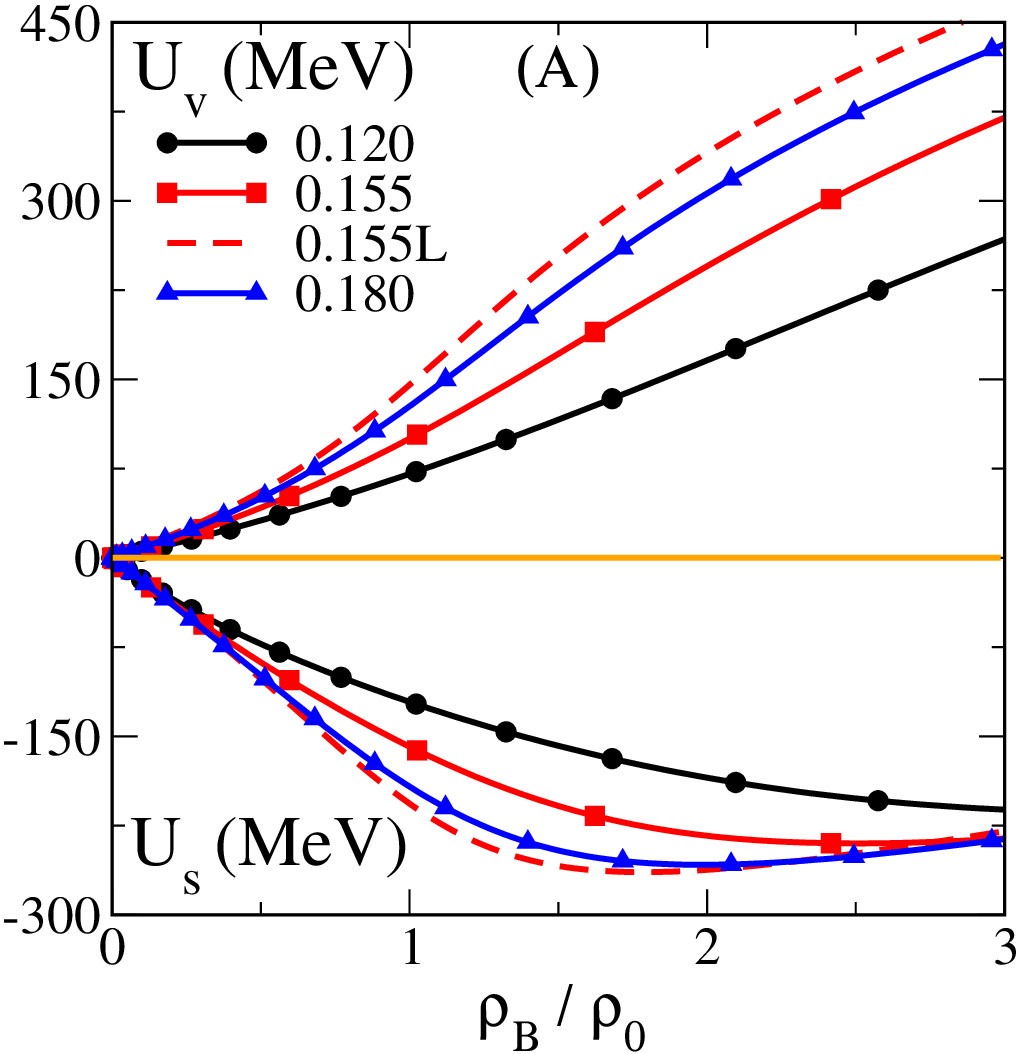}
\hskip 0.2in
\includegraphics[width=3.9cm]{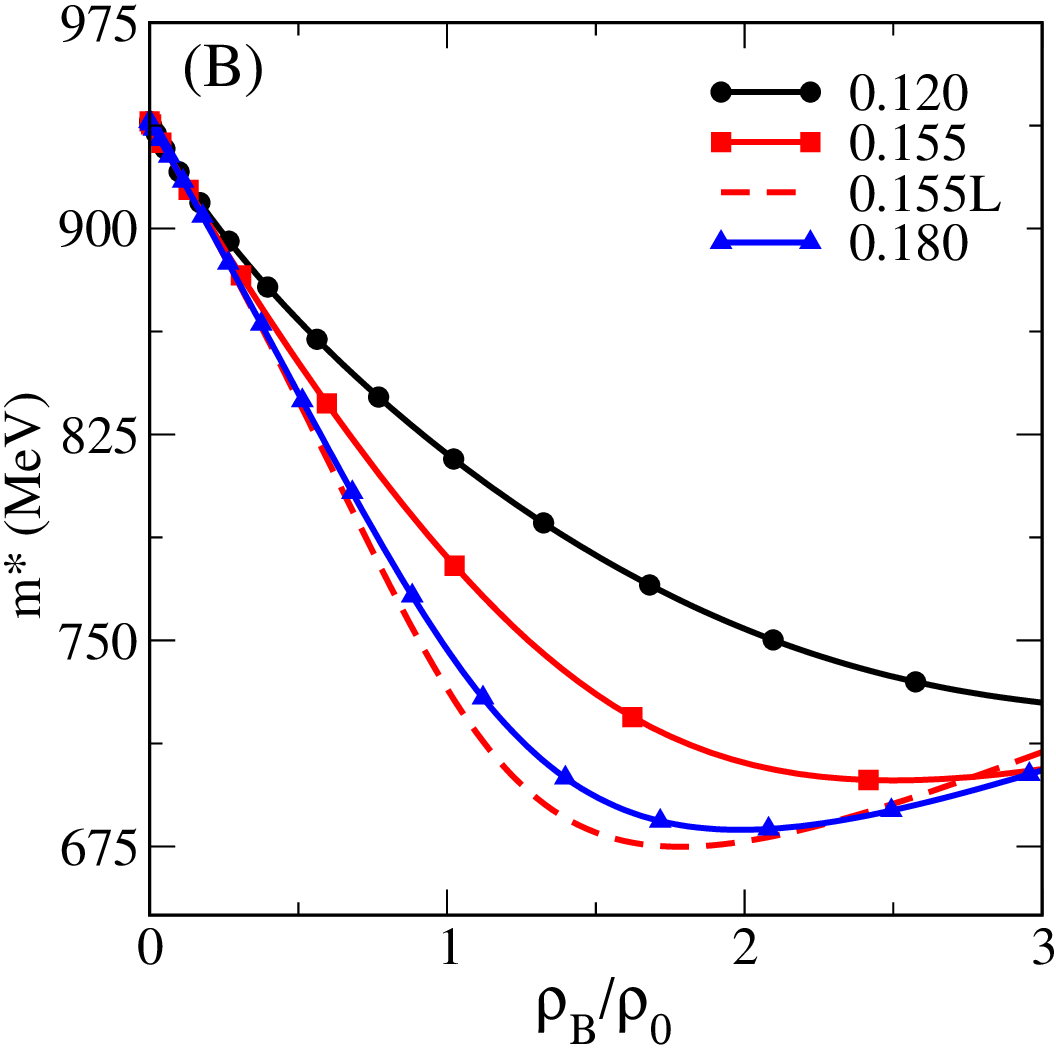}
\end{center}
\caption{(A) The vector ($U_v = g_{\omega}\omega_0$) and the scalar potentials ($U_s = g_{\sigma}\sigma_0$) for symmetric nuclear matter as a function of normalized baryon density for different $\rho_0$. The dashed curve corresponds to the linear interaction in the scalar field at $\rho_0 = 0.155 fm^{-3}$. (B) The nucleon effective mass is plotted as a function of normalized baryon density.}
\end{figure}

\begin{figure}[ht]
\vskip 0.2in
\begin{center}
\includegraphics[width=3.9cm]{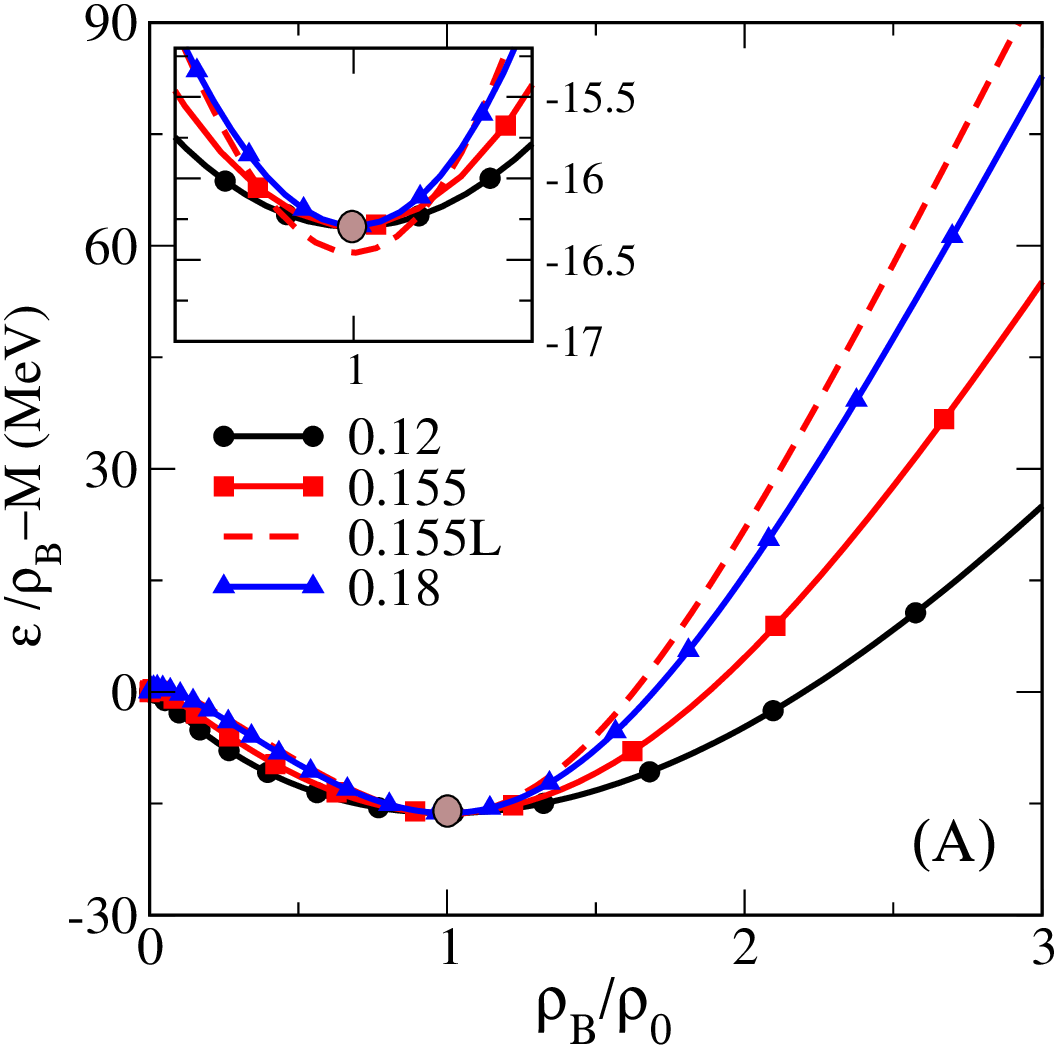}
\hskip 0.2in
\includegraphics[width=4cm]{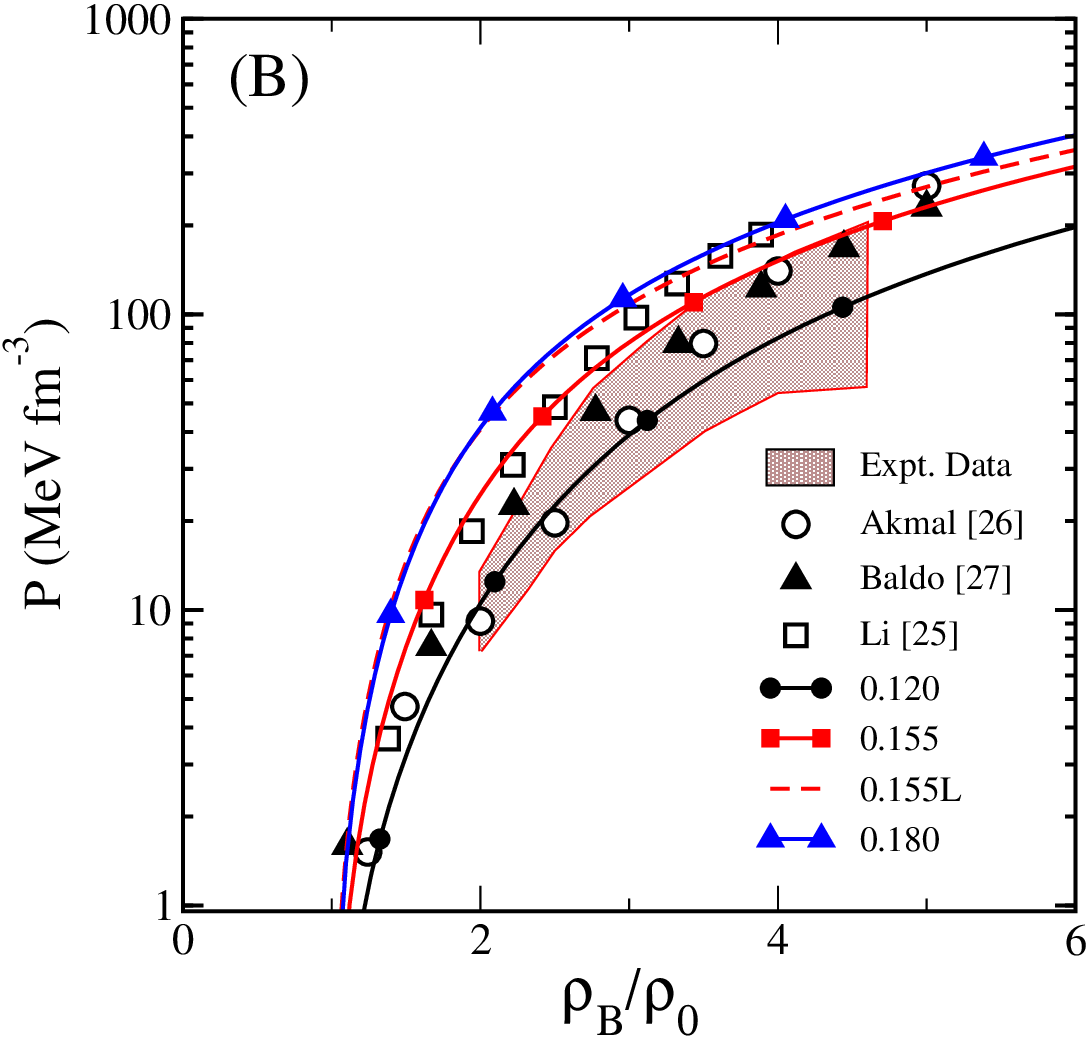}
\end{center}
\caption{(A) Binding energy per particle for symmetric nuclear matter plotted as a function of normalized baryon density. (B) Comparison of the present theoretical prediction with the HIC data for symmetric nuclear matter.}
\end{figure}

\begin{figure}[ht]
\begin{center}
\includegraphics[width=3.9cm]{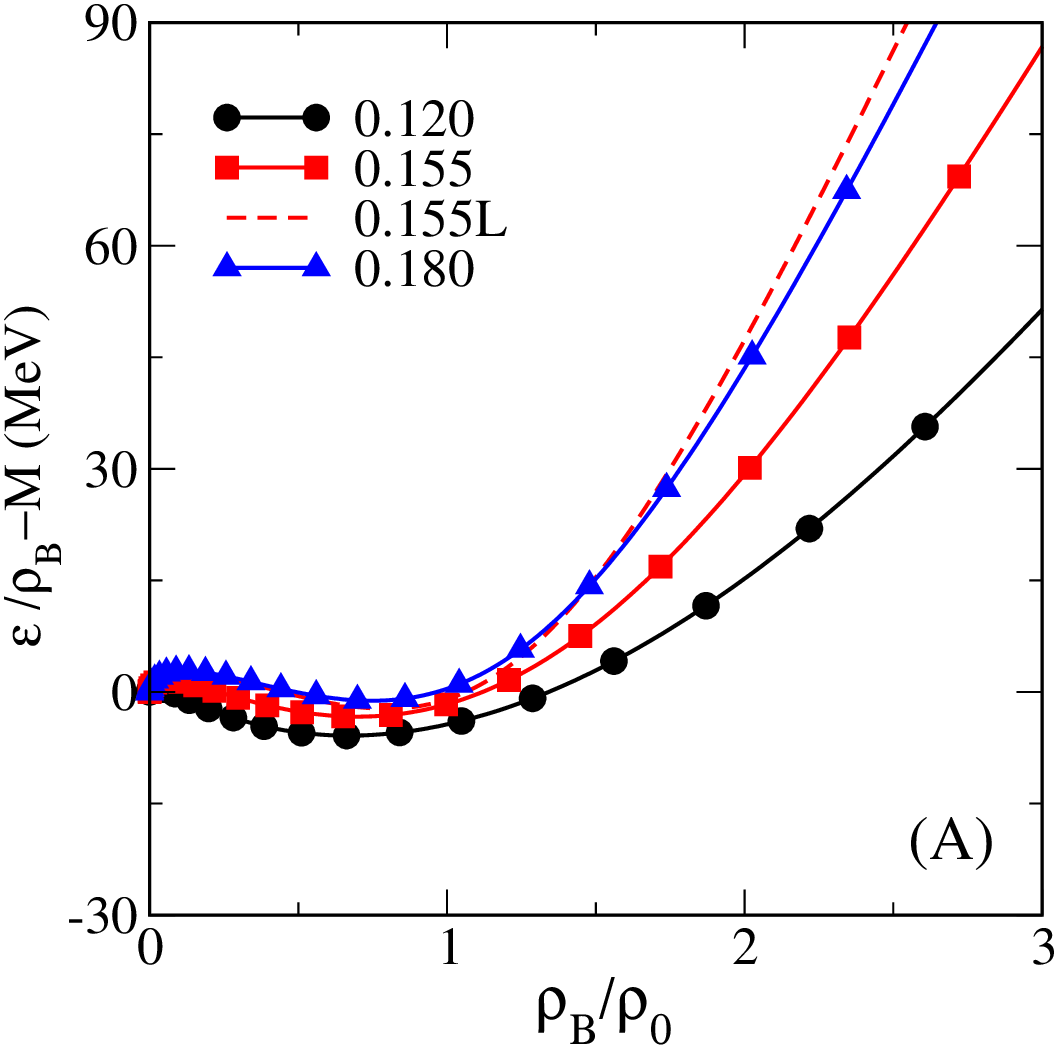}
\hskip 0.2in
\includegraphics[width=4cm]{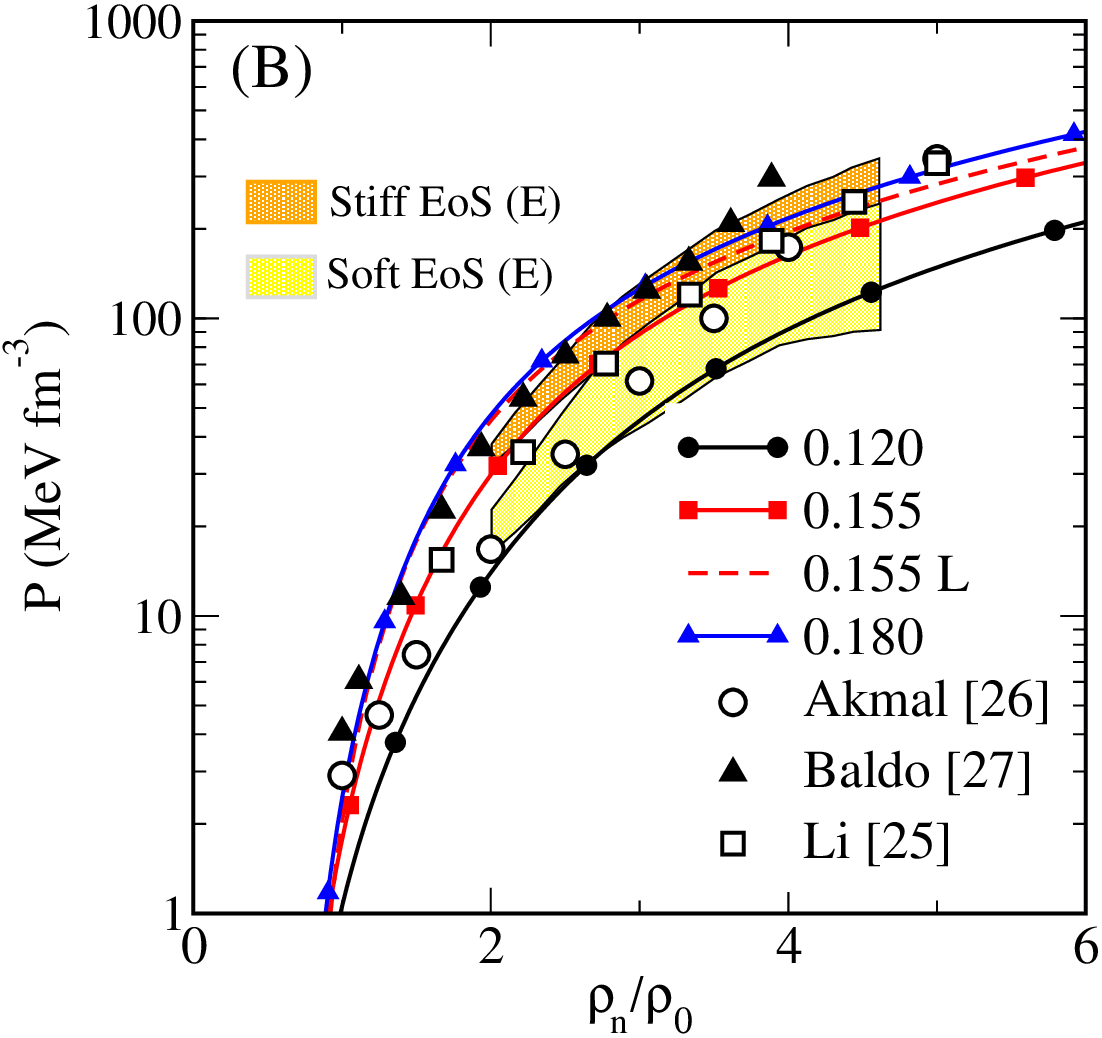}
\end{center}
\caption{(A) Binding energy per particle for pure neutron matter plotted as a function of normalized baryon density. (B) Comparison of the present theoretical prediction with the HIC data for pure neutron matter.}
\end{figure}

The obtained mass of the scalar meson is displayed in Fig. 1(F) and compared with the recent limit from the experimental search for the mass of the scalar meson, shown with the shaded region, corresponding to $m_{\sigma} = 513 \pm 32$ MeV \cite{mura02}. We find that the scalar meson mass obtained beyond $\rho_0 > 0.155 fm^{-3}$ and the mass of linear model does not agree with the experimental data. It is interesting to find that the combined experimental constraints from the HIC data for incompressibility of the matter and the scalar meson mass surprisingly agree with the upper limit on $\rho_0$. Our result for $m_{\sigma}=(541 \pm 10) MeV$ corresponding to $\rho_0 = 0.155 \pm 0.005 fm^{-3}$ are also compatible with the BES II fitted Breit-Wigner forms to the $\pi\pi$ interactions ($m_{\sigma}= 541 \pm 39 MeV$) \cite{besii} and the theoretical estimate from the analysis of the deuteron binding energy in the linear $\sigma$ model ($m_{\sigma} = 550 \pm 30 MeV$) \cite{deu}.

The elements contributing to the nuclear matter saturation and equation of state can be known from potential of the fields acting in nuclear matter. What is plotted in Fig. 2(A) is the scalar ($U_s = g_{\sigma} \sigma_0$) and the vector potentials ($U_v = g_{\omega}\omega_0$) for symmetric nuclear matter up to $3\rho_0$. The saturation of nuclear matter is realized through the interplay of the vector and the scalar field at $\rho_0$, till then both the potentials seems to increase with density. The potentials are stronger for matter that saturates at higher density than those which saturates at lower density. For example, the scalar and the vector potential for $\rho_0 = 0.12 fm^{-3}$ comes out to be $\approx -122~ \& ~71 ~MeV$ respectively, whereas for $\rho_0 = 0.18 fm^{-3}$, it is $\approx -193~ \& ~131 ~MeV$. The corresponding values for $\rho_0 = 0.155 fm^{-3}$ is $\approx -160~ \& ~99 ~MeV$ in Comparison to the linear potential ($0.155 L$) which has values $\approx -207~ \& ~147 ~MeV$ for the scalar and vector counterparts respectively. Quantitavely, when the non-linear terms in the scalar field interactions are included, we witness a drop of 22 \% and 33 \% in the scalar and vector potentials respectively at $\rho_0$. However the vector potential grows monotonously with density, whereas the scalar potential seems to saturate after $1.5 \rho_0$ for all the cases and the potential drops thereafter. Consequently at higher densities, repulsion dominates over attraction. 

An essential element for the success of relativistic phenomenology, the nucleon effective mass is shown in Fig. 2(B) as a function of normalized baryon density. In the nuclear medium the effective mass decreases continuously till about $2\rho_0$ and increases again, because of the dominant repulsive force at higher densities. As described earlier, the effective mass of the nucleon in nuclear matter experiences repulsion, which takes over attraction at high densities. The amount of increase/ decrease is found to be saturation density dependant. Further the contribution from the nonlinear field softens the sharp fall/ rise in the effective mass, owing to the drop in the field potentials.

In Fig. 3(A), the binding energy per particle is plotted as a function of normalized baryon density for $\rho_0 = 0.12,0.155~\&~0.18 fm^{-3}$ for symmetric nuclear matter (SNM). We find that both the linear model and EOS corresponding to high $\rho_0$ results in stiff EOS. In the inset, nice agreement between the nonlinear predictions of the model in the vicinity of saturation density. In Fig. 3(B), we compare corresponding EOS with the experimental bound from HIC \cite{data02} as well as few theoretical predictions. Here, we find that EOS corresponding to the best parameterization ($\rho_0 = 0.155 fm^{-3}$) of the present model calculation lies on the upper bound of the flow data, and compatible with the DBHF prediction \cite{li} at low density. For comparison, we also show the EOS obtained from variational calculations \cite{akmal} and EOS corresponding to BHF with AV14 potential plus 3BF from Baldo et.al. \cite{Baldo}. Similar comparison is made in case of pure neutron matter shown in Fig. 4(A) and 4(B), where again we find our results to be compatible with the DBHF calculations and seem to lie on the upper bound of the soft regime of the flow data.

The contribution  from the scalar ($\varepsilon_{\sigma}$), the vector ($\varepsilon_{\omega}$) and the kinetic energy ($\varepsilon_{k}$) to the total energy density for SNM is shown in Fig. 5 up to $3 \rho_0$. The respective contribution for the linear model corresponding to $\rho_0 = 0.155 fm^{-3}$ is also displayed. It can be seen that at all densities the majority of the contribution comes from the kinetic energy followed by the vector and the scalar contributions. However, in the vicinity of $\rho_0$ the contribution from the scalar sector increases rapidly and is more than the vector counterpart. We find that till $\approx 1.5 \rho_0$, attraction is dominant over repulsion. At $\rho_0$, for $\rho_0=0.155 fm^{-3}$, the kinetic energy of the nucleon contributes $\approx$ 86\% to the total energy density of the matter followed by the scalar contribution of $\approx 7 \%$ and the rest is by the vector counterpart. Irrespective of the saturation density, we find that the scalar field is dominant than the vector part up to about $1.5 \rho_0$, however the scalar contribution falls sharply thereafter.

\begin{figure}[ht]
\vskip 0.3in
\begin{center}
\includegraphics[width=8cm]{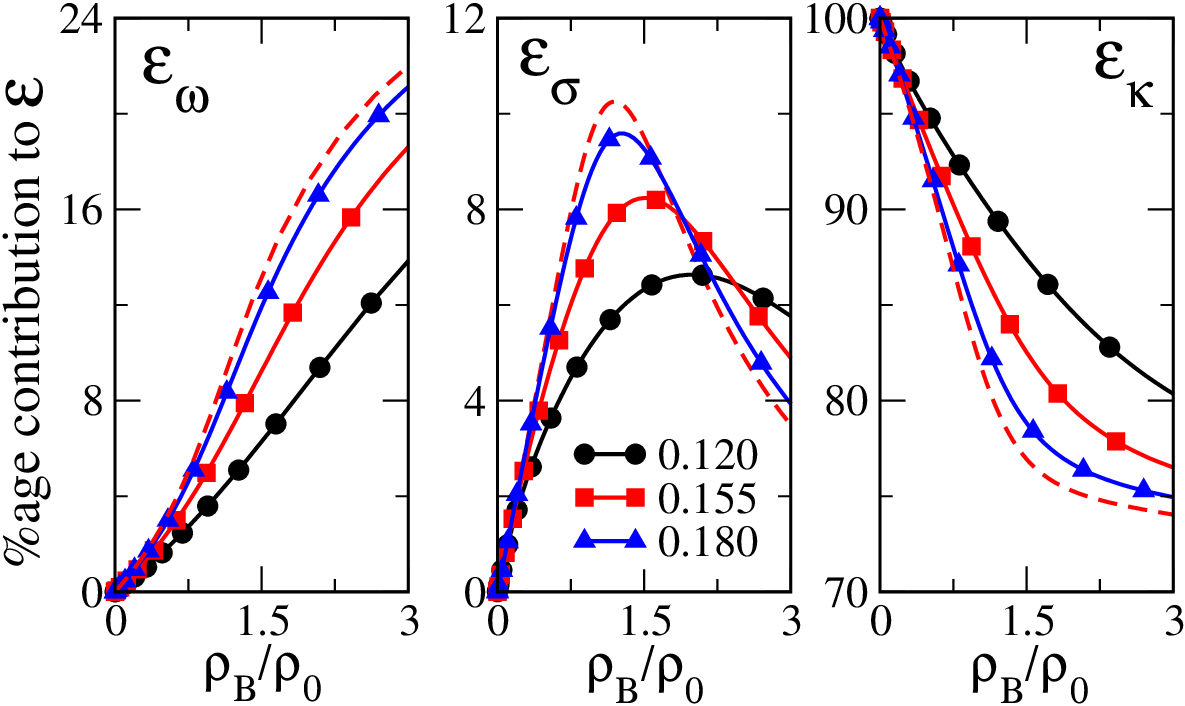}
\end{center}
\caption{The \% age contribution of the vector ($\varepsilon_{\omega}$), the scalar ($\varepsilon_{\sigma}$) and the kinetic energy of the nucleons ($\varepsilon_{k0}$) to the total energy density ($\varepsilon$) as a function of normalized baryon density for the specified EOS. The dashed curve corresponds to the linear model prediction for $\rho_0 = 0.155$.}
\end{figure}

\section{Summary and conclusions}

With a simple approach, we investigated the mechanism of saturation in nuclear matter and obtained parameters of the model to analyst the correlation between saturation properties and at different saturation density imposing chiral constraint. It is found that the nonlinearity in the scalar field provides freedom to fine tune the parameters so that they are compatible with the experimental value of the pion decay constant. However, the difference between the linear and the nonlinear scalar field interaction seems to wash out at higher saturation density. The present analysis projects the interlink between the fundamental properties of matter such as the nucleon effective mass, the nuclear incompressibility and the sigma meson mass. Aspects relating to the effect of nonlinearity in the scalar field interaction on the equation of state is investigated. The experimental value of the pion decay constant serves as a stern constraint on the vector coupling and in turn regulates the nucleon effective mass in the present model. Further the combined constraint from the HIC data \cite{data02} and sigma meson mass \cite{mura02} seems to provide the upper limit to $\rho_0 = 0.155 fm^{-3}$. Within an acceptable range $\rho_0 = 0.155 \pm 0.005$, the sigma meson mass is comes out in the range $m_{\sigma} = 546 \pm 10 MeV$, which agrees very well with recent experimental and theoretical bounds. Although within the specified range, the model predicts EOS with incompressibility $K \approx (354 - 398)$ and lies on the upper range of the bound from the HIC data, the EOS corresponding to $\rho_0 = 0.155 fm^{-3}$ is found to be compatible with the DBHF prediction \cite{li}. The resulting high value of nucleon effective mass in the present model is a consequence of the dominant repulsive force in matter at high density, which may have interesting implications in the astrophysical context such as the modelling of neutron stars where the matter density is speculated to be in the range $(3 - 10) \rho_0$ \cite{compact}. Overall, the model seems to provide a unified description of nuclear matter aspects, however the high incompressibility, although acceptable in the astrophysical domain needs to be brought down to enhance the applicability of the model in further studies.


\begin{thebibliography}{00}

\bibitem{wal74}J. D. Walecka, Ann. Phys.  83 (1974) 491; Phys. Lett. 79B (1978) 10.
\bibitem{boguta77} J. Boguta and A. R. Bodmer, Nucl. Phys. A  292 (1977) 413.
\bibitem{rmf} G. A. Lalazissis, S. Karatzikos, R. Fossion, D. Pena Artega, and P. Ring, Phys. Lett. B  671 (2009) 36; M. M. Sharma, Phys. Lett. B666 (2008) 140.
\bibitem{rmf1} B. G. Todd-Rutel and J. Piekarewicz, Phys. Rev. Lett 95 (2005) 122501.
\bibitem{rmf2} B. K. Sharma, P. K. Panda and S. K. Patra Phys. Rev.  C75 (2007) 035808.
\bibitem{ch01} M. Gell-Mann and M. Levy, Nuovo Cim. 16 (1960) 705.
\bibitem{le74} T. D. Lee and G. C. Wick, Phys. Rev. D9 (1974) 2291.
\bibitem{ch03} J. Boguta, Phys. Lett. B120 (1983) 34; Phys. Lett. B128 (1983) 19.
\bibitem{pdg} W. M. Yao et. al. (Particle Data Group), J. Phys. G33 (2006) 1 and 2007 partial update for the 2008 edition.
\bibitem{gl} N. K. Glendening, Ann. Phys. 168 (1986) 246.
\bibitem{five} N. K. Glendening, Nucl. Phys. A480 (1988) 597.
\bibitem{tkj08} T. K. Jha, H. Mishra and V. Sreekanth, Phys. Rev.  C77 (2008) 045801.
\bibitem{cons09} T. K. Jha and H. Mishra, Physical Review C78 (2008) 065802.
\bibitem{param1} B. M. Waldhauser, J. A. Maruhn, H. St\"ocker and W. Greiner, Phys. Rev. C38 (1988) 1003.
\bibitem{param2} K. C. Chung, C. S. Wang, A. J. Santiago and J. W. Zhang,
Eur. Phys. J. A12 (2001) 161.
\bibitem{data02} P. Danielewicz, R. Lacey and W. G. Lynch, Science 298 (2002) 1592.
\bibitem{mura02} H. Muramatsu et. al, Phys. Rev. Lett. 89 (2002) 251802.
\bibitem{nl3} G. A. Lalazissis, J. Konig, and P. Ring, Phys. Rev. C55 (1997) 540.
\bibitem{compact} N. K. Glendening, {\it Compact stars: Nuclear physics, particle physics, and general relativity}, Springer-Verlag, New York (2000).
\bibitem{nuclei} C. H. Johnson, D. J. Horen and C. Mahaux, Phys. Rev. C36 (1987) 2252.
\bibitem{furn98} R. J. Furnstahl, J. J. Rusnak and B. D. Serot, Nucl. Phys. A632 (1998) 607.
\bibitem{k1} H. M. M\"uller and B.D. Serot, Nucl. Phys. A606 (1996) 508.
\bibitem{besii} M. Ablikim et al. [BES II], Phys. Lett.  B598 (2004) 149.
\bibitem{deu} Yi-Bing Ding et al., J. Phys. G: Nucl. Part. Phys. 30 (2004) 841.
\bibitem{li} G. Q. Li, R. Machleidt, and R. Brockmann, Phys. Rev. C45 (1992) 2782.
\bibitem{akmal} A. Akmal, V. R. Pandharipande, and D. G. Ravenhall, Phys. Rev. C58 (1998) 1804.
\bibitem{Baldo} M. Baldo, I. Bombaci, and G. F. Burgio, Astron. Astrophys. 328 (1997) 274.


\end{thebibliography}
\end{document}